\documentclass[aps,prc,twocolumn,superscriptaddress,showpacs,nofootinbib]{revtex4-1}
\usepackage{graphicx,amsmath,amssymb,bm,color,physics,rotating}
\usepackage{epstopdf}

\usepackage{hyperref}
\usepackage{url}
\usepackage{lipsum}
\usepackage{cancel}
\hypersetup{
    colorlinks=true,
    linkcolor=blue,
    filecolor=magenta,      
    urlcolor=blue,
    citecolor=blue,
    }
\newcommand{\beq}{\begin{equation}}
\newcommand{\eeq}{\end{equation}}

\definecolor{joelred}{rgb}{0.81,0.13,0.16}

\definecolor{ryangreen}{rgb}{0.20,0.8,0.0}
\newcommand{\rc}[1]{\textcolor{black}{#1}}

\begin{document}

\title{Second-Order Perturbation Theory in Continuum Quantum Monte Carlo Calculations}

\author{Ryan Curry}
\affiliation{Department of Physics, University of Guelph, Guelph, Ontario N1G 2W1, Canada}
\author{Joel E. Lynn}
\affiliation{Institut f\"{u}r Kernphysik, Technische Universit\"{a}t Darmstadt, 64289 Darmstadt, Germany}
\affiliation{ExtreMe Matter Institute EMMI, GSI Helmholtzzentrum f\"ur
Schwerionenforschung GmbH, 64291 Darmstadt, Germany}
\author{Kevin E. Schmidt}
\affiliation{Department of Physics, Arizona State University, Tempe, Arizona 85287, USA}
\author{Alexandros Gezerlis}
\affiliation{Department of Physics, University of Guelph, Guelph, Ontario N1G 2W1, Canada}

\begin{abstract}
We report on the first results for the second-order perturbation theory correction to the ground-state energy of a nuclear many-body system in a continuum quantum Monte Carlo calculation. Second-order (and higher) perturbative corrections are notoriously difficult to compute in most \textit{ab~initio} many-body methods, where the focus is usually on obtaining the ground-state energy. By mapping our calculation of the second-order energy correction to an evolution in imaginary time using the diffusion Monte Carlo method, we are able to calculate these \rc{nuclear} corrections for the first time. After benchmarking our method in the few-body sector, we explore the effect of charge-independence breaking terms in the nuclear Hamiltonian.  We then employ the new approach to investigate the many-body, perturbative, order-by-order convergence that is fundamental in modern theories of the nucleon-nucleon interaction derived from chiral effective field theory. \rc{We find cutoff-dependent perturbativeness between potentials at higher chiral order and also that the difference between leading order (LO) and next-to-leading order (NLO) potentials is non-perturbative; both of these results have important implications for future nuclear many-body calculations.} Our approach is quite general and promises to be of wide applicability. 
\end{abstract}	
	

\maketitle

Nuclear many-body theory has been experiencing a renaissance over the last two decades, with several techniques now being routinely used to carry out first-principles studies of many strongly interacting nucleons \cite{Hergert:2020}. These include exact diagonalization approaches (when the basis size is manageable) \cite{RothNavratil:2007}, quantum
Monte Carlo QMC methods (stochastic evolution, whether in the continuum \cite{WiringaPieper:2002} or on a lattice \cite{Lee:2009}), more or less controlled truncations such as many-body perturbation theory (limiting oneself to the first several orders), \cite{Tichai_etal:2016,Hu:2016} self-consistent Green's function approaches (implementing a specific class of diagrams up to infinite order), \cite{Soma:2020} Coupled Cluster (generating $n$p$n$h excitations off a reference state) \cite{Hagen_etal:2014} and its younger cousin the in-medium Similarity Renormalization group \cite{Hergert_etal:2016}.  Several of these approaches are also very important in broader applications of the quantum many-body problem, including, for example, condensed matter physics and the study of ultracold atoms \cite{GiorginiPitaevskiiStringari:2008, BartlettMusial:2007, KolorencMitas:2011,Drut:2013,Shi:2021}. The difference between nuclear and other many-body techniques is driven by the two- (or higher-) body interaction at play.

Using nucleons as degrees of freedom, the interaction between particles has historically been written down as the combination of short-range/contact terms and terms corresponding to the exchange of mesons (most notably pions). The study of nuclear interactions itself has been rejuvenated by the aforementioned renaissance in nuclear many-body techniques (see, e.g., Refs.~\cite{Hebeler:2010,Gezerlis:2013,Coraggio:2013,Hagen:2014,Gezerlis:2014,Carbone:2014,Roggero:2014,Wlazlowski:2014,Soma:2014,
Elhatisari:2015,Tews:2016,Piarulli:2018,Lonardoni:2018,Marcucci:2019,Schiavilla:2019,Carbone:2019,Tichai:2020,Zhang:2020,Yang:2021,Sobczyk:2021,Chen:2022}); another driving force has been the recasting of earlier phenomenological forces in the language of effective field theory \cite{HammerKonidvanKolck:2020}. The contact terms arising in the context of pionless effective field theory (EFT), also of applicability to cold atoms, where the range of the interaction is much shorter than the interparticle spacing, are then supplemented by the exchange of the lightest mesons, most notably in the context of chiral EFT. Several open questions remain at the forefront of the study of nuclear forces, such as the few-body observables that should be used as constraints, the importance of higher orders and uncertainty estimates, as well as the determination of an appropriate (renormalization-group invariant) power-counting scheme \cite{Tews_etal:2022}. Once again, much of the recent progress (and hope for further developments) has been driven by the interplay of novel nuclear many-body techniques and designer nuclear forces.

As noted, an important open question, on which much work has been dedicated recently, relates to the relative importance of power counting and nonperturbative renormalization in chiral EFT. Given the significance of the interface between nuclear forces and many-body techniques, a newcomer might be forgiven for assuming that the problem of perturbative vs nonperturbative force vs many-body technique will have received considerable attention. However, until very recently this has not been the case; to see why, let us recall the quantum-mechanical expressions for 1st and 2nd order corrections to the ground-state energy for a given perturbation $V^{\prime}$:
\begin{align}
E_0^{(1)} &= \bra{\psi_0^{(0)}}V^{\prime}\ket{\psi_0^{(0)}} \nonumber
\\
E_0^{(2)} &= -\sum_{k \neq 0} \frac{ \left| \bra{\psi_0^{(0)}}V^{\prime}\ket{\psi_k^{(0)}} \right|^2}{E_k^{(0)}-E_0^{(0)}} \label{2ndorder}
\end{align}
where $\psi_0$, $\psi_k$, $E_0$, and $E_k$ are the ground and $k$th excited eigenstates and energies of the unperturbed Hamiltonian, respectively; crucially, all of these are many-body entities, i.e., in contradistinction to the textbook case, they cannot be tackled analytically.  The problem becomes immediately apparent: While the first-order correction $E_0^{(1)}$ is easily computable in a many-body context (e.g., in the diffusion Monte Carlo (DMC) method the propagator would involve the unperturbed Hamiltonian and the observables would involve $V^{\prime}$), computing the second-order correction $E_0^{(2)}$ is dramatically more complicated, as it requires knowledge of the complete energy spectrum $E_k$. While progress has been made on tackling individual excited states in an \textit{ab~initio} setting, e.g., by imposing the appropriate symmetries on a variational wave function \cite{PieperWiringaCarlson:2004}, that is still a manual process (separately designing a new wave function for each new state) that cannot be easily generalized beyond a handful of cases.

This significant task is woefully underdiscussed in the many-body literature; even the few exceptions cast the problem in quite different language (cumulant autocorrelation functions in \cite{Caffarel_Hess:1991}, auxiliary fields in \cite{Lu_etal:2022}, or static response in \cite{Gaudoin:2010}).  In the present Letter, we show how to map the computation of the second-order energy correction $E_0^{(2)}$ to an evolution in imaginary time; we do this (for the first time) in the context of a continuum QMC technique oriented toward nuclear forces, but the recasting we employ is much more general and could plausibly be implemented in other areas of physics or distinct many-body techniques. Both for reasons of clarity and in order to establish our approach's credentials, we start from \rc{(i)} a trivial (two-body) application as a benchmarking exercise; we then turn to \rc{(ii)} a more challenging application, resulting from the introduction of a charge-independence breaking term in the nuclear interaction.  Significantly, we then turn to the main application, namely \rc{(iii)} using the new approach as a detailed probe of the perturbativeness (or lack thereof) of chiral EFT interactions at the many-body level. 

Crucially, the nonperturbative calculation does not correspond to an \textit{ad~hoc} interaction employed merely to make higher-order terms small \cite{Wlazlowski:2014, Lu_etal:2022} but to an actual chiral EFT interaction in widespread use (the local next-to-next-to leading order (N$^2$LO) interaction of Gezerlis et al \cite{Gezerlis:2014}). 
\rc{While Ref.~\cite{Lu_etal:2022} pioneered the use of second-order perturbation theory in nuclear-physics quantum Monte Carlo, it was limited to the use of a single cutoff. 
In other words, our new approach will allow us to study how perturbative different nuclear interactions are at the many-body level, going beyond first-order perturbation theory.}
 The fact that we address three distinct applications reflects the generality of the proposed methodology.

We compute Eq. (\ref{2ndorder}) using the DMC method. We first start with many sets of particle positions (walkers) that are distributed according to a trial wave function $\psi_T$ by a preliminary variational Monte Carlo (VMC) calculation \cite{Ceperley:1997}. The DMC method then projects out the lowest energy eigenstate $\psi_0$ by treating the Schr\"odinger equation as a diffusion equation in imaginary time $\tau$ and propagating the trial wave function up to large $\tau$. This can be demonstrated by expanding the trial wave function in terms of the complete set of exact eigenstates, $\psi_T = \sum_i \alpha_i\psi_i^{(0)}$, and then applying the imaginary time propagation operator,
\begin{align}
   \lim_{\tau\to\infty}\psi(\tau) &=
   \lim_{\tau\to\infty}\text{exp}[-(\hat{H}_0-E_T)\tau]\psi_T \propto\psi_0^{(0)}. \label{dmc1}
\end{align}

Since DMC projects out the ground-state wave function of the system of interest, we start our calculation of the second-order energy correction by assuming we have access to the ground state, and consider the quantity,
\begin{align}
\label{CompleteI}
I(\mathcal{T}) = \int_0^{\mathcal{T}} d\tau \bra{\psi_0^{(0)}} V^{\prime}e^{-[\hat{H}_0 - E_0^{(0)}]\tau}V^{\prime}\ket{\psi_0^{(0)}},
\end{align}
By an insertion of the identity, and then a splitting of the sum between $k=0$ and $k \neq 0$, Eq.~(\ref{CompleteI}) can be recast as
\begin{align} \label{I_fitform}
I(\mathcal{T}) &= (E_0^{(1)})^2 \mathcal{T} - \nonumber
\\
&\sum_{k \neq 0}^{\infty} \frac{\left |\bra{\psi_k^{(0)}}V^{\prime}\ket{\psi_0^{(0)}} \right|^2}{E_k^{(0)} - E_0^{(0)}} \left[e^{-[E_k^{(0)}-E_0^{(0)}]\mathcal{T}} - 1 \right].
\end{align}
In the limit of long imaginary time the exponential in the second term can be neglected and by comparing with Eq.~(\ref{2ndorder}), we see we have recovered both the first- and second-order corrections to the ground-state energy,
\begin{align}
I(\mathcal{T \rightarrow \infty})  = (E_0^{(1)})^2 \mathcal{T} - E_0^{(2)}.
\label{eq:eq5}
\end{align}
Therefore, for a given perturbation $V^{\prime}$, the second-order correction to the ground-state energy can be computed in a many-body context by implementing Eq.~(\ref{CompleteI}). 

\begin{figure}[t]
\centering
\includegraphics[width=0.45\textwidth]{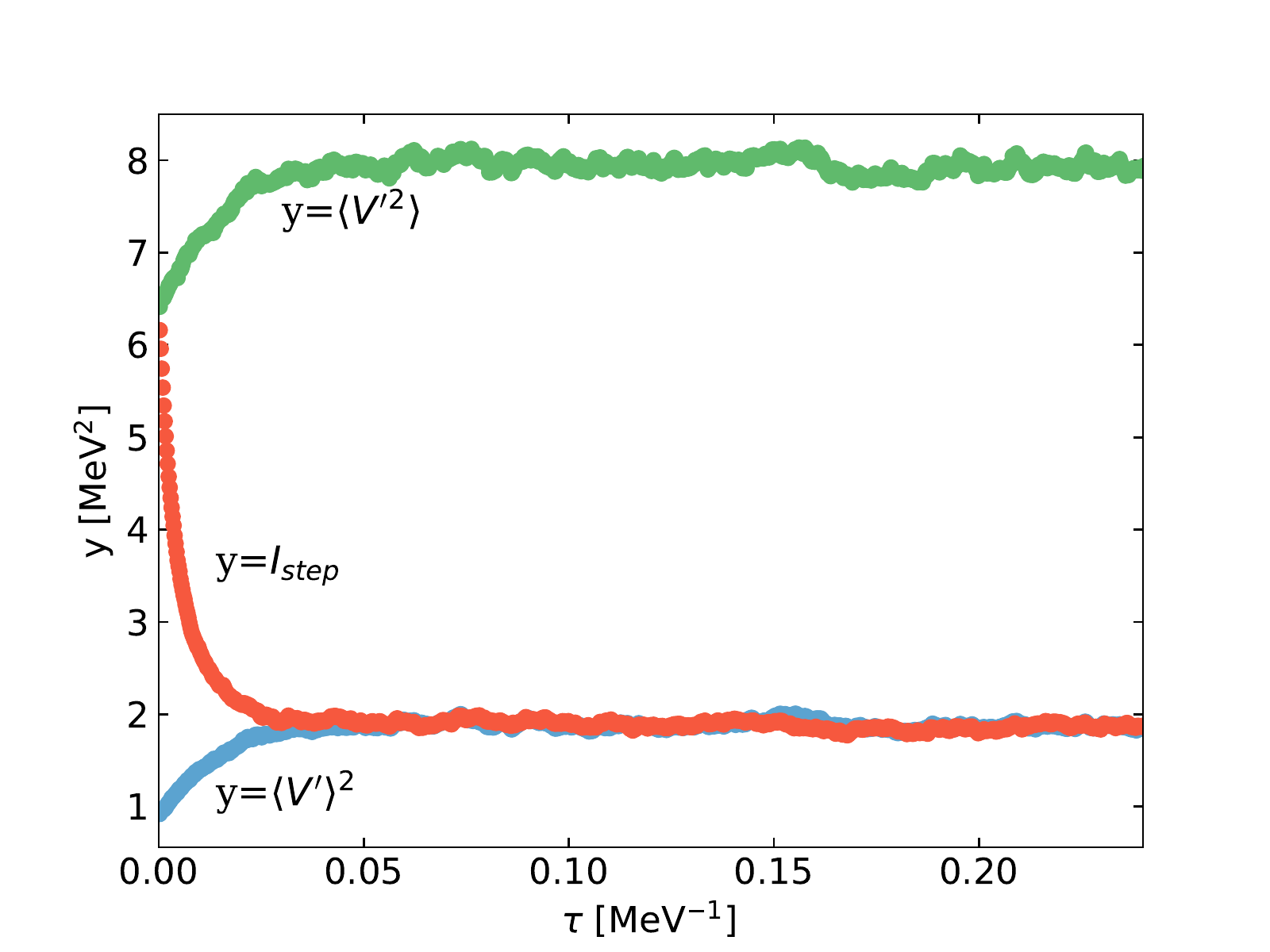} 
   \caption{A DMC calculation of Eq.(\ref{I_0}) for perturbation from a neutron-neutron interaction to a neutron-proton interaction. By comparing with Eq.(\ref{I_0}) we see the expected behavior. At very small $\tau$ we essentially recover $\left< V^{\prime 2} \right>$ and in the limit of long $\tau$ we have the expected result that $I_\text{step}$ decays down to $\left< V^{\prime} \right>^2$. (See text for more details.)}
\label{fig.time_evo}
\end{figure}

The question then, is how to evaluate Eq.~(\ref{CompleteI}) in a QMC context, where we are starting with a realistic trial wave function, and not the ground state. The plan will be to compute the integrand of Eq.~(\ref{CompleteI}), fit its behavior to the form of Eq.~(\ref{I_fitform}), and finally employ an extrapolated estimate. In a DMC calculation the full imaginary time $\tau$ is broken into many successive small-$\tau$ segments, which means our quantity of interest can be written as
\begin{align}
\bra{\psi_0^{(0)}} V^{\prime}\prod_n e^{-[\hat{H}_0-E_0]\Delta\tau}V^{\prime}\ket{\psi_0^{(0)}},
\end{align}
or as a Monte Carlo integral,
\begin{align} \label{final_I0}
I_{\text{step}}= \int d\boldsymbol{R} d\boldsymbol{R}^{\prime}V^{\prime}(\boldsymbol{R}^{\prime})\tilde{G}(\boldsymbol{R}^{\prime},\boldsymbol{R};\Delta\tau)V^{\prime}(\boldsymbol{R})\psi^2_T(\boldsymbol{R}),
\end{align}
where $\psi_T$ is our trial wave function and $\tilde{G}(\boldsymbol{R}^{\prime},\boldsymbol{R};\Delta \tau)$ is the importance-sampled short-imaginary-time propagator \cite{Pudliner_etal:1997}. We can then finally rewrite Eq.~(\ref{final_I0}) as a twice importance sampled Monte Carlo integral,
\begin{align}
\label{I_0}
\rc{I_{\text{step}} \approx  \frac{\sum_{i=0}^{\mathcal{N}-1} w_i V^{\prime}(\boldsymbol{R}_i^{\prime})V^{\prime}(\boldsymbol{R}_{i,0})}{\sum_{i=0}^{\mathcal{N}}w_i}},
\end{align}
\rc{where $\mathcal{N}$ is the number of walkers, the weights $w_i$ are calculated according to $\text{exp}[-\Delta \tau (E_L(\boldsymbol{R})+E_L(\boldsymbol{R}^{\prime}) - 2E_T)/2]$ as is standard \cite{Foulkes:2001}}. 
The $\boldsymbol{R}_i^{\prime}$ are positions distributed according to $\tilde{G}$ and the $\boldsymbol{R}_{i,0}$ are positions distributed according to the trial wave function. An example calculation of Eq. (\ref{CompleteI}) in a DMC context by way of Eq. (\ref{I_0})is shown in Fig.~\ref{fig.time_evo} (the physical details will be introduced in due course). By comparing with Eq.~(\ref{I_0}) we can trace the behavior of the integral $I$. We should expect that at very small $\tau$ the integral should be equal to $\left<V^{\prime 2}\right>$, since the exponential in Eq.~(\ref{CompleteI}) has only just started to decay. In addition, we expect the integral to decay to $\left<V^{\prime}\right>^2$ in the limit of large-$\tau$ based on Eq.~(\ref{I_fitform}).

Therefore, to compute the second-order energy correction using Eq.~(\ref{CompleteI}), we compute Eq.~(\ref{I_0}) at every step of the DMC process. To find an accurate estimate for the second-order correction, we perform two DMC runs. First we perform an initial run starting from walkers distributed according to the trial wavefunction, and then a second run starting from walkers distributed according to the projected ground-state wave function.  We then compute an extrapolated estimate between this ground-state initialized run and the initial trial wavefunction run, which is necessary for the expectation value of any operator that does not commute with the Hamiltonian when forward walking techniques are not employed \cite{Gezerlis_Carlson:2010}.

To benchmark this new method, the second-order energy corrections were calculated and compared against nonperturbative results for a variety of few-body systems in a harmonic oscillator potential with oscillator frequency $\omega$. To perform these calculations we used a harmonic oscillator eigenstate basis trial wave function, with variational parameters differing slightly from the ground-state wave function of trapped particles, in the spirit of Ref.~\cite{ChangBertsch:2007}.

Our initial few body tests involve starting from noninteracting particles in the harmonic trap, and perturbatively including an interaction in the form of a simple Gaussian interaction, $V^{\prime}=a e^{-q^2 (\boldsymbol{r}_2 - \boldsymbol{r}_1)^2}$, where $a=1.0 ~\hbar \omega$ and $q=1.0 ~\text{fm}^{-1}$. We first carry this out for simple two-particle systems in a variety of harmonic trap strengths. As was to be expected, we see that the quality of the answer provided by perturbation theory (whether 1st or 2nd order) is reasonable, as long as the perturbation is small. (See the first three rows of Table~\ref{table:1}.) Though we produce the answer using the DMC machinery, for few-body systems the results can also be produced quasi-analytically.

\begin{table}[t]
\centering
\begin{tabular}{c c c c c c} 
 \hline
 \hline
 $N$ & $\hbar \omega$ [MeV] &$E^{[0]}$ & $E^{[1]}$ & $E^{[2]}$ & Non-Pt. \\ [1.0ex] 
 \hline
 2 & 1  & 3.0008(1) & 3.1951(2) & 3.1769(3) & 3.1768(2)\\ 
 2 & 2  & 3.0006(2) & 3.1790(2) & 3.1728(2) & 3.1713(2)\\ 
 2 & 5 & 3.0013(2) & 3.1231(2) & 3.1220(2) & 3.1211(1)\\ 
 \hline
 \hline
  2 & 1 & 2.57335(7) & 2.58457(8) & 2.58437(8) & 2.58427(6)\\ 
  4 & 1  & 6.5582(4) & 6.5876(4) & 6.5865(4) & 6.5866(4)\\ 
  6 & 1  & 10.0465(4) & 10.0898(4) & 10.0885(6) & 10.0876(4)\\ 
 \hline
 \hline
\end{tabular}
\caption{Summary of our second-order energy corrections for few-body systems compared against nonperturbative results. $E^{[n]}$ refers to the sum of energy corrections up to the $n$th order. All energies are in units of $\hbar \omega$. Upper: Results for a Gaussian perturbation between two noninteracting particles, Lower: Results for a perturbation from the neutron-neutron interaction to the neutron-proton interaction. }
\label{table:1}
\end{table}

However, because our interest lies in applying our method to nuclear systems, we now introduce our general unperturbed Hamiltonian
\begin{align} \label{H}
\hat{H}_0 = \sum_{i=1}^N \left(-\frac{\hbar^2}{2m}  \nabla_i^2 + \frac{1}{2} \omega^2 r_i^2 \right) + \sum_{i<j^{\prime}}^{N}   V(r_{ij^{\prime}}),
\end{align}
where $N$ is the total number of particles and $V(r_{ij^{\prime}})$ is the interaction between spin up (primed) and spin down (unprimed) particles. In the case of the nucleon-nucleon interaction, this $V(r_{ij^{\prime}})$ contains contributions like the tensor
force, spin-orbit, etc. Here, with a view to abstracting away the details of the interaction
(allowing us to employ sophisticated wave functions in DMC) we limit ourselves to 
low density/$s$-wave interactions. 

Historically, nuclear many-body calculations have used phenomenological potentials that are fit to experimental data~\cite{WiringaStoksSchiavilla:1995, WiringaPieper:2002}. For this initial nuclear investigation we employ a phenomenological potential of the P\"oschl-Teller type as in Ref.~\cite{Gezerlis_Carlson:2010},
\begin{align} \label{poschlteller}
V(r) = - v_0 \frac{ \hbar^2}{m} \frac{\mu^2}{\cosh^2{\mu r}}, 
\end{align}
where $v_0$ and $\mu$ are parameters that can be tuned in order to reproduce the scattering length and effective range of a nuclear interaction, which describes the interaction completely at low energies~\cite{Bethe:1949}. 

\begin{figure}[t]
\centering
\includegraphics[width=0.45\textwidth]{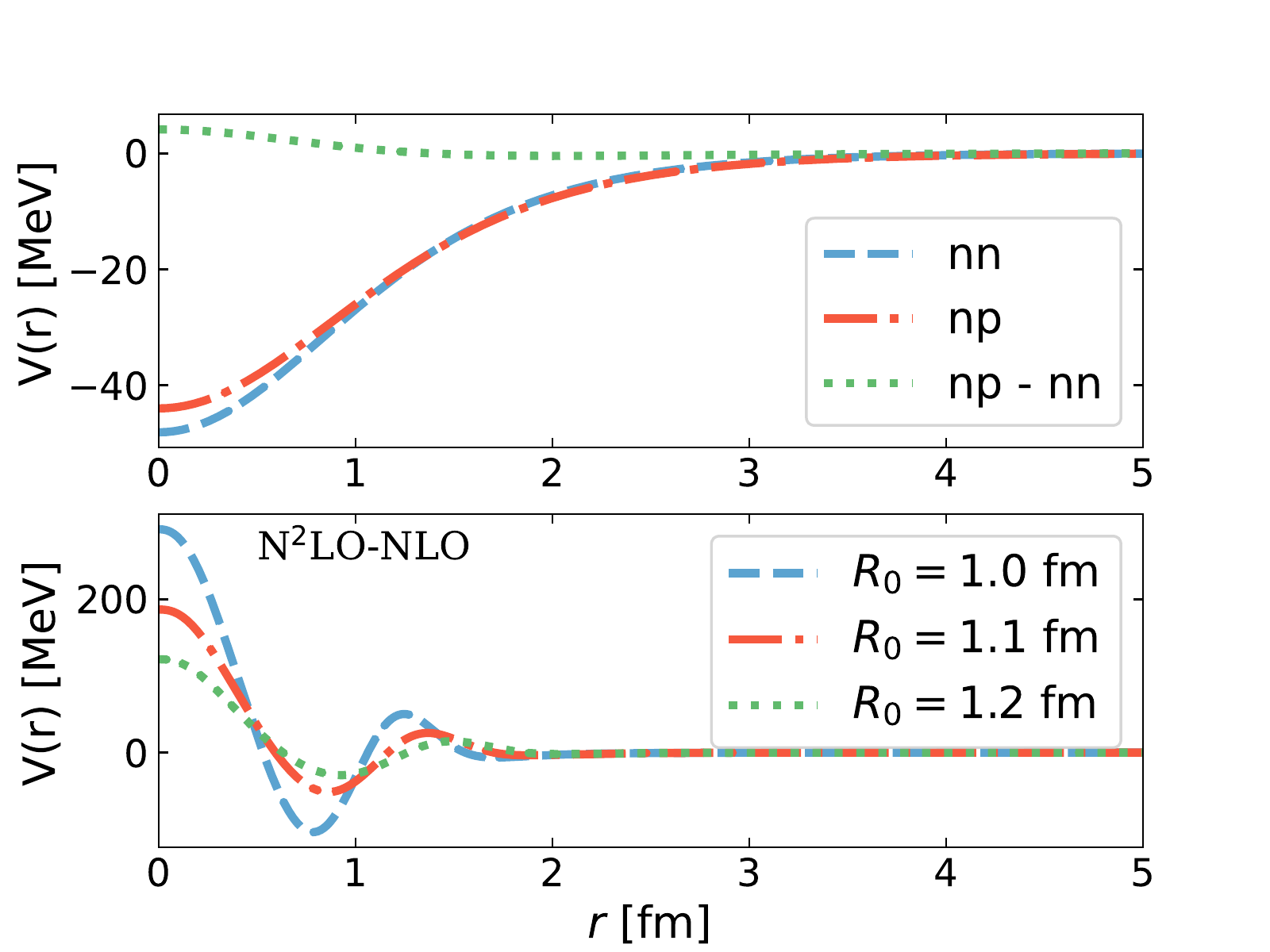} 
   \caption{Upper panel: Perturbing potential given by the difference
   between the $s$-wave neutron-neutron interaction and the
   neutron-proton interaction. Lower panel: Perturbation given by the
   difference between the $^1$S$_0$ N$^2$LO and NLO chiral EFT
   interactions for a range of coordinate space cutoff values.}
\label{fig.potentials}
\end{figure}

Having benchmarked the method at the two-body level, we now consider a range of particle numbers up to $N=6$, and set our unperturbed interaction to correspond to a neutron-neutron interaction
(still in a harmonic trap). Starting from this unperturbed neutron system, we apply a perturbation such that we move the system from a two-neutron interaction to a neutron-proton one \rc{(a charge-independence breaking term that is known to be of small magnitude)}
\begin{align}
V^{\prime} = V_{np} - V_{nn},
\end{align}
as illustrated in the upper panel of Fig.~\ref{fig.potentials}.

The nonperturbative ground-state energies of particles interacting through the $nn$ or $np$ potentials are compared against our perturbative calculations in the bottom three rows of Table~\ref{table:1}. We can see that while the first-order correction does a reasonable job of approximating the neutron-proton case, the second-order correction is required in order to accurately reproduce the nonperturbative neutron-proton ground-state energy to within one standard deviation.

Having applied our method to the few-body sector where the perturbation is not too difficult to handle, we now turn to a realistic nuclear system. For this study we modify our Hamiltonian from Eq. (\ref{H}) removing the external one-body oscillator potential. We also use $N=66$ particles as this gives a very good approximation to the thermodynamic limit~\cite{PalkanoglouDiakonosGezerlis:2020}. In addition, since both the VMC and the DMC methods depend strongly on the choice of trial wave function \rc{(the latter due to the presence of the fermion sign problem, tackled via the fixed-node prescription)}, and we are interested in the physics of pure infinite neutron matter, where pairing is known to be important~\cite{Gezerlis_Carlson:2010,Dean_HjorthJensen:2003}, we make use of periodic boundary conditions and the Jastrow-BCS wave function \cite{Carlson_etal:2003}, 

\begin{align}
\psi_T = \prod_{i \neq j^{\prime}}f(r_{ij^{\prime}}) \mathcal{A} \left[ \prod_{i<j^{\prime}}\phi(r_{ij^{\prime}}) \right].
\end{align}
As in Eq.~(\ref{H}), the primed indices correspond to spin up neutrons, and the unprimed indices to spin down neutrons. The pairing functions $\phi(\mathbf{r}) = \tilde{\beta}(r) + \sum_{\mathbf{n}} \alpha_n e^{i \mathbf{k}_{\mathbf{n}} \cdot \mathbf{r}}$ capture both long- and short-range physics as described in Ref.~\cite{Carlson_etal:2003}.

\begin{figure}[t]
\vspace*{-5mm}
\centering
\includegraphics[width=0.45\textwidth]{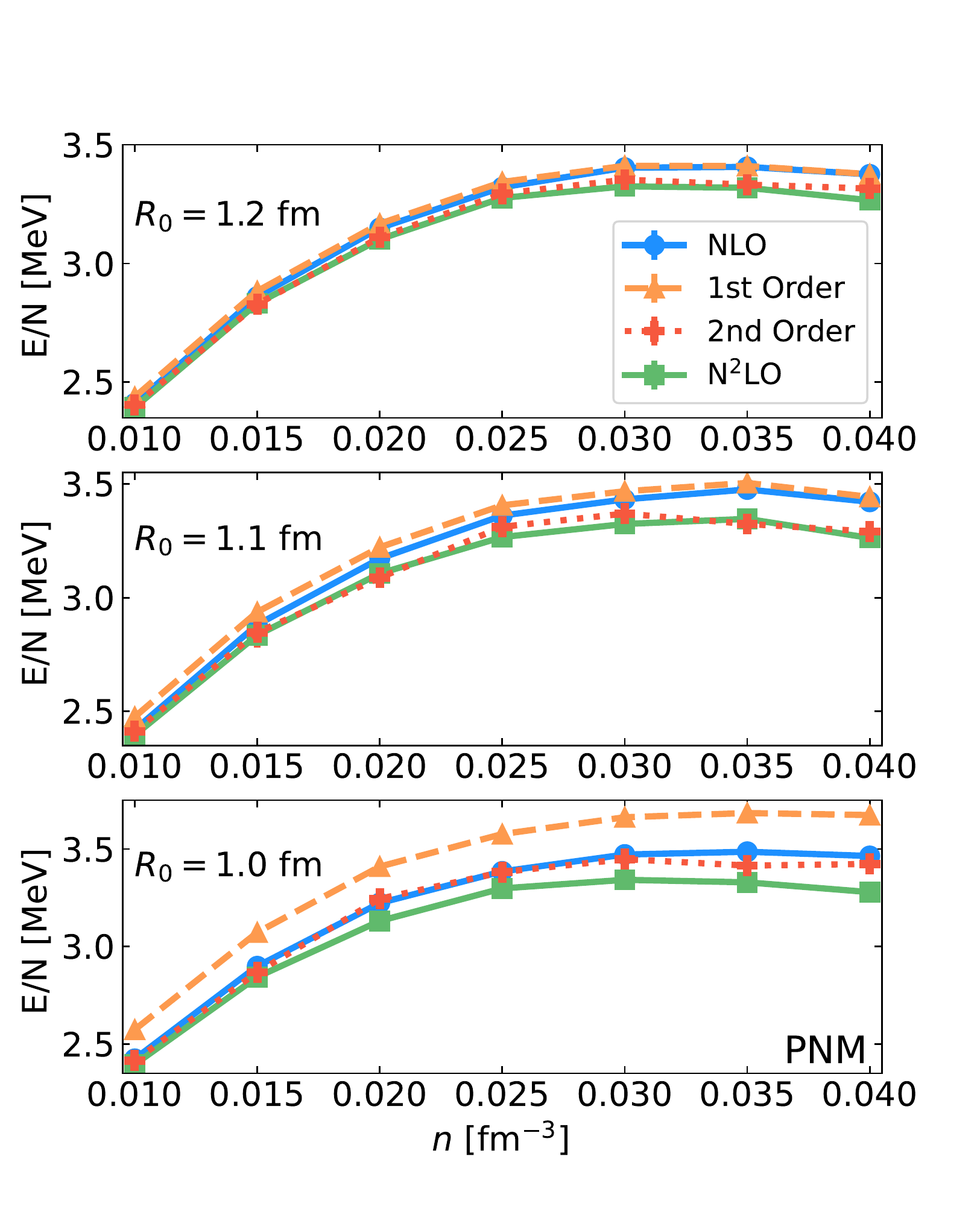}
\vspace*{-5mm}
\caption{Comparison between the nonperturbative results for NLO, N$^2$LO and the perturbative results for \rc{a perturbation} from NLO to N$^2$LO at both first and second order \rc{for pure neutron matter (PNM) with $N=66$ over a range of system densities and coordinate-space cutoffs.} \label{fig.results1}}
\end{figure}

In recent decades, the \textit{ab initio} many-body community has largely moved on from phenomenological potentials like the ones used for our preliminary tests, and modern interactions have been built using a chiral EFT framework. In addition to reproducing known two-body experimental results, these chiral EFT interactions also capture the important symmetries from the underlying theory of quantum chromodynamics~\cite{MachleidtEntem:2011}. Chiral EFT also provides a systematic expansion for the nuclear force, with terms included at successive orders based on their importance following a power-counting scheme~\cite{EpelbaumHammerMeiBner:2009},
\begin{align}
   V_\text{chiral} = V^{(0)} + V^{(2)} + V^{(3)} + \dots,
\end{align}
where the superscript keeps track of the power of the expansion parameter $Q/\Lambda_b$, which depends on the momenta of the nucleons, or the pion mass ($Q$), and the scale at which the chiral EFT expansion breaks down ($\Lambda_b$). \rc{This is a \textit{perturbative} expansion so it should, properly speaking, be employed as such in a many-body technique; as stated earlier, such a fusing of non-perturbative and perturbative approaches has been hampered due to technical difficulties in the past; having proposed our new method, we are now in a position to tackle this issue directly.}

 This expansion structure informed the choice of interactions used for this work. To study the physics of neutron matter, we use the chiral EFT interactions of Ref.~\cite{Gezerlis:2013} that have been tuned to reproduce the dominant $^1$S$_0$ channel interaction between opposite spin neutrons. We start from an unperturbed system with an interaction containing terms up to NLO and perturb to include terms at N$^2$LO
\rc{\begin{equation}
V^{\prime} = V_{\text{N$^2$LO}} - V_{\text{NLO}},
\end{equation}} 
\noindent for a range of coordinate space cutoffs, as seen in the bottom panel of Fig.~\ref{fig.potentials}.

The result of carrying out such a many-body computation up to second-order in perturbation theory is shown in Fig.~\ref{fig.results1} for a number of densities; \rc{this is the main upshot of the present work}. \rc{(Finite-size effects are quite minimal for a homogeneous gas~\cite{Gaudoin:2010,Gezerlis:2021}.)} It is worth emphasizing that when the core of the potential is soft ($R_0 \geq 1.1\text{ fm}$) the second-order perturbation theory results for treating the difference between N$^2$LO and NLO as a perturbation match the nonperturbative N$^2$LO results to within 1\% error.

This follows the general trend found via an exact diagonalization for the deuteron in Ref.~\cite{Lynn_etal:2014}. In each case we see that the first-order correction is positive, and the second-order correction is negative. It can also be seen that at the higher densities the agreement between the N$^2$LO results and the second-order perturbation theory results is worsened. This is likely due to the fact that at large densities other interaction channels become important and so we are no longer in the regime where the neutron-neutron interaction is correctly described by a pure $^1S_0$ interaction. 

The results for $R_0 = 1.0$ fm in Fig.~\ref{fig.results1} suggest that the strong repulsive core coincides with the need to use at least third-order corrections, and raises the question as to whether the perturbative-based chiral EFT interactions are appropriately behaved \cite{Nogga_etal:2005, Long_Yang:2012}.   We have also attempted to perturb from NLO to LO, with a perturbation $V_\text{LO} - V_\text{NLO}$, (and vice versa) with limited success. This provides further evidence that the difference between LO and NLO is not perturbative.
 
In summary, we have developed a method for calculating the second-order perturbation theory correction to the ground-state energy in a continuum quantum Monte Carlo context. We have benchmarked this method against nonperturbative results in the few-body sector, and performed initial calculations for few-body neutron systems in a trap that could pave the way for more complicated four-species calculations in the future.  Finally, we have completed the first calculations for pure infinite neutron matter that include the second-order correction to the ground-state energy. In addition, we have also tested the behavior of popular chiral EFT interactions and found indications of nonperturbativeness. In the future, this method has broad applicability and could become a staple in the toolbox of anyone making use of DMC methods or its extensions (e.g. the auxiliary-field diffusion Monte Carlo method or the Green's Function Monte Carlo method. 

The work of R.C. and A.G. was supported by the Natural Sciences and Engineering Research Council (NSERC) of Canada and the
Canada Foundation for Innovation (CFI). 
The work of J.E.L. was supported by the ERC Grant No. 307986 STRONGINT and the BMBF under Contract No. 05P15RDFN1.
Computational resources were provided by SHARCNET and NERSC.


\end{document}